\documentclass[preprint,12pt]{elsarticle} 
\usepackage{graphicx} 
\usepackage{amssymb} 
\usepackage{amsmath} 

\journal{Physics Letters B} 

\def\lamb#1#2{$^{#1}_{\Lambda}${#2}}

\begin{document} 

\begin{frontmatter} 

\title{Is there a bound \lamb{3}{n}?} 

\author[a]{Avraham Gal\corref{cor1}} 
\cortext[cor1]{Corresponding author: Avraham Gal, avragal@vms.huji.ac.il} 
\author[b]{Humberto Garcilazo} 
\address[a]{Racah Institute of Physics, The Hebrew University, 91904 
Jerusalem, Israel} 
\address[b]{Escuela Superior de F\' \i sica y Matem\'aticas \\ 
Instituto Polit\'ecnico Nacional, Edificio 9, 07738 M\'exico D.F., Mexico} 

\date{\today} 

\begin{abstract} 
The HypHI Collaboration at GSI argued recently for a \lamb{3}{n} 
($\Lambda nn$) bound state from the observation of its two-body $t$+$\pi^-$ 
weak-decay mode. We derive constraints from several hypernuclear systems, in 
particular from the $A=4$ hypernuclei with full consideration of $\Lambda N
\leftrightarrow\Sigma N$ coupling, to rule out a bound \lamb{3}{n}. 
 
\end{abstract} 


\end{frontmatter}

\section{Introduction} 
\label{sec:intro} 

The lightest established $\Lambda$ hypernucleus is known since the early days 
of hypernuclear physics to be \lamb{3}{H}$^{T=0}$, in which the $\Lambda$ 
hyperon is weakly bound to the $T=0$ deuteron core, with ground-state (g.s.) 
separation energy $B_{\Lambda}$(\lamb{3}{H})=0.13$\pm$0.05~MeV and spin-parity 
$J^P$=$\frac{1}{2}^+$. There is no evidence for a bound spin-flip partner with 
$J^P$=$\frac{3}{2}^+$. For a brief review on related results deduced from past 
emulsion studies of light hypernuclei, see Ref.~\cite{Davis05}. 

As for \lamb{3}{H}$^{T=1}$, given the very weak binding of the $\Lambda$ 
hyperon in the $T=0$ g.s., and that the $T=1$ $NN$ system is unbound, 
it is unlikely to be particle stable against decay to $\Lambda+p+n$. 
Similarly, assuming charge independence, $\Lambda nn$ is not expected to 
be particle stable. As early as 1959 just six years following the discovery 
of the first $\Lambda$ hypernucleus, it was concluded by Downs and Dalitz 
upon performing variational calculations of both $T=0,1$ $\Lambda NN$ 
systems that the isotriplet (\lamb{3}{n}, \lamb{3}{H}$^{T=1}$, \lamb{3}{He}) 
hypernuclei do not form bound states \cite{DD59}. This issue was revisited 
in Refs.~\cite{Gar87,MKGS95,Hiyama14} using various versions of Nijmegen 
hyperon-nucleon ($YN$) potentials within $\Lambda NN$ Faddeev equations for 
states with total orbital angular momentum $L=0$ and all possible values of 
total angular momentum $J$ and isospin $T$. Again, no $\Lambda nn$ bound state 
was found in any of these studies as long as \lamb{3}{H}$^{T=0}$($J^P=\frac
{1}{2}^+$) was only slightly bound. Similar conclusions were reached in 
Refs.~\cite{Gar07a,Gar07b,Gar14} based on chiral constituent quark model 
$YN$ interactions, and in Ref.~\cite{Gazda14} based on recently constructed 
NLO chiral EFT $YN$ interactions~\cite{Haiden13}. Note that $\Lambda N
\leftrightarrow\Sigma N$ coupling was fully implemented in the more recent 
\lamb{3}{n} studies \cite{MKGS95,Hiyama14,Gar07a,Gar07b,Gar14,Gazda14}. 
A more general discussion of stability vs. instability for \lamb{3}{n} in 
the context of neutral hypernuclei with strangeness $-$1 and $-$2 has been 
given very recently in Ref.~\cite{JMR14}. 

A claim for particle stability of \lamb{3}{n} has been made recently by the 
HypHI Collaboration \cite{Rap13} observing a signal in the $t+\pi^-$ invariant 
mass distribution following the bombardment of a fixed graphite target by 
$^6$Li projectiles at 2$A$ GeV in the GSI laboratory. The binding energy 
of the conjectured weakly decaying \lamb{3}{n} is 0.5$\pm$1.1$\pm$2.2~MeV, 
with a large standard deviation $\sigma$=5.4$\pm$1.4~MeV. As noted above, 
there is unanimous theoretical consensus based on $\Lambda NN$ bound-state 
calculations that \lamb{3}{n} cannot be particle stable. However, possible 
connections to other hypernuclear systems, in particular the $A=4$ bound 
isodoublet hypernuclei (\lamb{4}{H}, \lamb{4}{He}), need to be explored. 
The present work addresses this issue by establishing connections that make 
it clear why a bound \lamb{3}{n} cannot be accommodated into hypernuclear 
physics. Assuming charge-symmetric $\Lambda N$ interactions, 
$V_{\Lambda p}=V_{\Lambda n}$, we demonstrate some unacceptable implications 
of a bound \lamb{3}{n} to $\Lambda p$ scattering in Sect.~\ref{sec:Lp}, 
and to \lamb{3}{H}$^{T=0}$ in Sect.~\ref{sec:Lpn}. Consequences of $A=4$ 
hypernuclear spectroscopy with full consideration of charge-symmetric 
$\Lambda N\leftrightarrow\Sigma N$ couplings are derived for \lamb{3}{n} 
in Sect.~\ref{sec:Lpnn} by applying methods that differ from those used 
in the combined analysis of $A=3$ and $A=4$ hypernuclei by Hiyama et 
al.~\cite{Hiyama14}, reaffirming that \lamb{3}{n} is unbound. Our results 
are discussed and summarized in Sect.~\ref{sec:summ}, with additional remarks 
made on the possible role of charge-symmetry breaking (CSB) and $\Lambda NN$ 
interaction three-body effects, concluding that a bound \lamb{3}{n} 
interpretation of the $t+\pi^-$ signal in the HypHI experiment is outside 
the scope of present-day hypernuclear physics.

\section{\lamb{3}{n} vs. $\Lambda p$ scattering} 
\label{sec:Lp} 

To make a straightforward connection between the low-energy $\Lambda N$ 
scattering parameters and the three-body $\Lambda NN$ system we follow the 
method of Ref.~\cite{Gar87} in solving $YNN$ Faddeev equations with two-body 
$YN$ input pairwise separable interactions constructed directly from given 
low-energy $YN$ scattering parameters. For simplicity we neglect in this 
section the spin dependence of the low-energy $\Lambda N$ scattering 
parameters, setting $a_s=a_t$ for the scattering length and $r_s=r_t$ with 
values $r$=2.5 or 3.5~fm for the effective range, spanning thereby a range 
of values commensurate with most theoretical models and also with the analysis 
of measured $\Lambda p$ cross sections at low energies \cite{Alex68}. 
By using Yamaguchi form factors within rank-one separable interactions, 
we then compute critical values of scattering length $a$ required to bind 
successively the $T=0$ and $T=1$ $\Lambda NN$ systems, with results shown 
in Table~\ref{tab:acrit}. 

\begin{table}[hbt]
\begin{center}
\caption{Values of the spin-independent $\Lambda N$ scattering length 
$a$ required to bind $T=0$ and $T=1$ $\Lambda NN$ states as indicated, 
for two representative values of the spin-independent effective range 
$r$, and calculated values of the $\Lambda p$ total cross section at 
$p_{\Lambda}$=145~MeV/c. The measured value at the lowest momentum bin 
available is $\sigma_{\Lambda p}^{\rm tot}(p_{\Lambda}$=145$\pm$25~MeV/c)=180$
\pm$22~mb \cite{Alex68}. Calculated values of $B_{\Lambda}$(\lamb{3}{H}$
^{T=0}$) are listed in the last column for $\Lambda N$ interactions that just 
bind \lamb{3}{n}, in contrast to $B_{\Lambda}^{\rm exp}$(\lamb{3}{H})=0.13$\pm
$0.05~MeV.} 
\begin{tabular}{cccccccc} 
\hline 
 & \multicolumn{2}{c}{$B_{\Lambda}^{T=0}$=0} & \multicolumn{2}{c}
{$B_{\Lambda}^{T=0}$=0.13 MeV} & \multicolumn{3}{c}{$B_{\Lambda}^{T=1}$=0 
(\lamb{3}{n} just bound)} \\  
\hline 
$r$ & $a$ & $\sigma_{\Lambda p}^{\rm tot}$ & $a$ & 
$\sigma_{\Lambda p}^{\rm tot}$ & $a$ & $\sigma_{\Lambda p}^{\rm tot}$ &  
$B_{\Lambda}^{T=0}$ \\ 
(fm) & (fm) & (mb) & (fm) & (mb) & (fm) & (mb) & (MeV) \\
\hline 
2.5 & $-$1.185 & 129.7 & $-$1.498 & 192.5 & $-$4.491 & 953.8 & 2.59 \\ 
3.5 & $-$1.405 & 152.4 & $-$1.895 & 239.7 & $-$5.930 & 943.1 & 1.74 \\ 
\hline
\end{tabular} 
\label{tab:acrit} 
\end{center} 
\end{table} 

Exceptionally large values of $\Lambda N$ scattering lengths are seen to be 
required to bind \lamb{3}{n}, and the low-energy $\Lambda p$ cross sections 
thereby implied exceed substantially the measured cross sections as shown by 
the $\Lambda N$ cross sections evaluated at the lowest momentum bin reported 
in Ref.~\cite{Alex68}. Of the three $B_{\Lambda}$ values tested in the table, 
only $B_{\Lambda}^{T=0}$=0.13~MeV is consistent with the reported $\Lambda p$ 
cross sections, including their uncertainties. In the last column of the table 
we also listed the $\Lambda$ separation energies in \lamb{3}{H} that result 
once \lamb{3}{n} has just been brought to bind. These calculated values are 
much too big to be reconciled with $B_{\Lambda}^{\rm exp}$(\lamb{3}{H})=0.13$
\pm$0.05~MeV.

\section{\lamb{3}{n} vs. \lamb{3}{H}} 
\label{sec:Lpn} 

The \lamb{3}{n} vs. \lamb{3}{H} discussion in this section is limited to using 
$s$-wave $\Lambda N$ effective interactions, providing a straightforward 
extension of earlier studies~\cite{DD59,Gar87}. Effects of possibly 
substantial $\Lambda N \leftrightarrow \Sigma N$ coupling, as generated by 
strong one-pion exchange in Nijmegen meson-exchange potentials \cite{Rijken10} 
and in recent chirally based potentials \cite{Haiden13}, are discussed in 
Sect.~\ref{sec:Lpnn}. 

\begin{table}[!ht]
\begin{center}
\caption{$\Lambda$ separation energies $B_{\Lambda}$(\lamb{3}{H}$^{T=0}$) 
(in MeV) calculated for both $J^P=\frac{1}{2}^+,\frac{3}{2}^+$, using $\Lambda 
N$ separable interactions based on the low-energy parameters Eq.~(\ref{eq:C'}) 
with $V_t$ multiplied by a factor $x$ up to values allowing \lamb{3}{n} to 
become bound, as indicated by following the values of its Fredholm determinant 
(FD) at $E=0$.} 
\begin{tabular}{cccc} 
\hline 
$x$ & \lamb{3}{n} FD($E=0$) & 
$B_{\Lambda}$[\lamb{3}{H}$^{T=0}(\frac{1}{2}^+)$] & 
$B_{\Lambda}$[\lamb{3}{H}$^{T=0}(\frac{3}{2}^+)$] \\ 
\hline  
1.00 & 0.55 & 0.096 & unbound \\ 
1.10 & 0.47 & 0.147 & 0.124 \\ 
1.20 & 0.39 & 0.211 & 0.448 \\ 
1.30 & 0.31 & 0.288 & 0.986 \\ 
1.40 & 0.21 & 0.381 & 1.704 \\ 
1.50 & 0.12 & 0.488 & 2.598 \\ 
1.60 & $+$0.015 & 0.612 & 3.659 \\ 
1.61 & $+$0.004 & 0.625 & 3.772 \\ 
1.62 & $-$0.006 & 0.638 & 3.890 \\  
\hline
\end{tabular} 
\label{tab:Fad} 
\end{center} 
\end{table} 

Following Ref.~\cite{Gar87} we solve Faddeev equations for \lamb{3}{n} and 
\lamb{3}{H} using simple Yamaguchi separable $s$-wave interactions fitted 
to prescribed input values of singlet and triplet scattering lengths $a$ 
and effective ranges $r$, thereby relaxing the spin-independence assumption 
of the preceding section. Of the four Nijmegen interaction models A,B,C,D 
studied there, only C reproduces the observed binding energy of \lamb{3}{H}, 
binding also the $\frac{3}{2}^+$ spin-flip excited state just 11 keV above 
the $\frac{1}{2}^+$ g.s. To get rid of this excited state, we have slightly 
changed the input parameters of model C. In this model, denoted C', the input 
$\Lambda N$ low-energy parameters are (in fm): 
\begin{equation} 
a_s=-2.03, \;\;\; r_s=3.66, \;\;\;\;\; a_t=-1.39, \;\;\; r_t=3.32\;.  
\label{eq:C'} 
\end{equation} 
The \lamb{3}{H}$^{T=0}$($J^P$=$\frac{1}{2}^+$,$\frac{3}{2}^+$) separation 
energies obtained by solving the appropriate $\Lambda NN$ Faddeev equations 
are listed in Table~\ref{tab:Fad}. The row marked $x=1$ corresponds to using 
$\Lambda N$ interaction based on the low-energy parameters Eq.~(\ref{eq:C'}), 
and subsequent rows correspond to multiplying the $\Lambda N$ triplet 
interaction $V_t$ by $x>1$ in order to bind \lamb{3}{n} (\lamb{3}{H}$^{T=1}$). 

Inspection of Table~\ref{tab:Fad} shows that while the $\Lambda$ separation 
energies increase upon varying $x$, a by-product of this increase is that 
\lamb{3}{H}$^{T=0}(\frac{3}{2}^+)$ quickly overtakes \lamb{3}{H}$^{T=0}
(\frac{1}{2}^+)$ becoming \lamb{3}{H} g.s. This is understood by observing 
that the weights with which $V_t$ and the singlet interaction $V_s$ enter 
a simple folding expression for the $\Lambda$--core interaction are given by 
\begin{equation} 
J^P=\frac{1}{2}^+: \; (T+\frac{1}{2})\;V_t+(\frac{3}{2}-T)\;V_s, 
\;\;\;\;\;\;\;\; J^P=\frac{3}{2}^+: \; 2V_t, 
\label{eq:folding} 
\end{equation} 
so that $V_t$ is the only $\Lambda N$-interaction component affecting 
\lamb{3}{H}$^{T=0}(\frac{3}{2}^+)$ besides being more effective in 
binding \lamb{3}{n} than binding \lamb{3}{H}$^{T=0}(\frac{1}{2}^+)$. 
Subsequently, beginning with $x=1.614$, \lamb{3}{n} becomes bound as 
indicated by the corresponding Fredholm determinant at $E=0$ going through 
zero. Note that the $(2J+1)$-averaged $B_{\Lambda}^{T=0}$(\lamb{3}{H}) is then 
$\approx$2.76~MeV, in rough agreement with the spin-independent analysis of 
the previous section (cf. first row in Table~\ref{tab:acrit}). Similar results 
are obtained when replacing the parameters (\ref{eq:C'}) of model C' by 
those of model C, used in Ref.~\cite{Gar87}, and repeating the  procedure 
summarized in Table~\ref{tab:Fad}. A bound \lamb{3}{n} is therefore in strong 
disagreement with the binding energy $B_{\Lambda}^{\rm exp}$(\lamb{3}{H})=
0.13$\pm$0.05~MeV determined for \lamb{3}{H}$_{\rm g.s.}$ and with its 
spin-parity $J^P=\frac{1}{2}^+$.

\section{\lamb{3}{n} vs. \lamb{4}{H}} 
\label{sec:Lpnn} 

$\Lambda N\leftrightarrow\Sigma N$ coupling cannot be ignored in quantitative 
calculations of $\Lambda$ hypernuclear binding energies. One-pion exchange 
induces a strong coupling in the $YN$ $^3S_1-{^3D_1}$ channel which dominates 
the effective $V_t$ contribution in \lamb{3}{H} three-body calculations, 
independently of whether using NSC97-related $YN$ interactions as 
in Refs.~\cite{MKGS95,Hiyama14} or NLO chiral $YN$ interactions in 
Ref.~\cite{Nogga12}. In the $YN$ $^1S_0$ channel, in contrast, $\Lambda N
\leftrightarrow\Sigma N$ coupling is weak. Here we employ $G$-matrix 
$0s_N0s_{Y}$ effective interactions devised by Akaishi et al.~\cite{Akaishi00} 
from the Nijmegen soft-core interaction model NSC97 and used in binding 
energy calculations of the $A=4,5$ $\Lambda$ hypernuclei. Of particular 
significance in the present context is the $\approx$1.1~MeV splitting of the 
$0^+_{\rm g.s.}$--$1^+_{\rm exc}$ spin-doublet levels in the isodoublet 
hypernuclei \lamb{4}{H}--\lamb{4}{He} which cannot be reconciled with theory 
without substantial $\Lambda N\leftrightarrow\Sigma N$ contribution. These 
$0s_N0s_{Y}$ effective interactions were extended by Millener to the $p$ shell 
and tested there successfully in a comprehensive analysis of hypernuclear 
$\gamma$-ray measurements \cite{Millener12}. For a recent application to 
neutron-rich hypernuclei, see Ref.~\cite{Galmil13}. The $0s_N0s_{Y}$ $\Lambda 
N\leftrightarrow\Sigma N$ effective interaction $V_{\Lambda\Sigma}$ assumes 
a spin-dependent central interaction form 
\begin{equation} 
V_{\Lambda\Sigma}=({\bar V}_{\Lambda\Sigma}+\Delta_{\Lambda\Sigma}{\vec s}_N
\cdot{\vec s}_{Y})\sqrt{4/3}\;{\vec t}_N\cdot{\vec t}_{\Lambda\Sigma}, 
\label{eq:V_YN} 
\end{equation}     
where ${\vec t}_{\Lambda\Sigma}$ converts a $\Lambda$ to $\Sigma$ in isospace, 
with matrix elements 
\begin{equation} 
{\bar V}_{\Lambda\Sigma}=2.96 (3.35)~{\rm MeV}, \;\;\;\;\; 
\Delta_{\Lambda\Sigma}=5.09 (5.76)~{\rm MeV}   
\label{eq:VbarDelta} 
\end{equation} 
derived from the Nijmegen model version NSC97e (NSC97f) as given in 
Ref.~\cite{Galmil13} (Ref.~\cite{Millener08}). As for the diagonal 
$0s_N0s_{Y}$ interactions, we will constrain the spin-dependent $\Lambda N$ 
interaction $\Delta_{\Lambda\Lambda}$ matrix elements by fitting, together 
with ${\bar V}_{\Lambda\Sigma}$ and $\Delta_{\Lambda\Sigma}$, to the 
excitation spectrum of the $A=4$ hypernuclei. Finally, the detailed 
properties of the $\Sigma N$ interaction hardly matter in view of the 
large energy denominators of order $M_{\Sigma}-M_{\Lambda}\approx 80$~MeV 
with which they appear. The binding-energy contribution arising from 
$V_{\Lambda\Sigma}$ is then given to a good approximation schematically 
by $|\langle V_{\Lambda\Sigma} \rangle |^2/80$ (in MeV). 

The nonvanishing matrix elements of the spin-independent term in 
Eq.~(\ref{eq:V_YN}) are given in closed form by 
\begin{equation} 
\langle (J_N T,s_{\Sigma}t_{\Sigma})JT|V_{\Lambda\Sigma}
(\Delta_{\Lambda\Sigma}=0)|(J_N T,s_{\Lambda}t_{\Lambda})JT \rangle
=\sqrt{4T(T+1)/3}\;{\bar V}_{\Lambda\Sigma}, 
\label{eq:V_LS} 
\end{equation} 
where $s_{\Sigma}=s_{\Lambda}=\frac{1}{2}$, $t_{\Sigma}=1$, $t_{\Lambda}=0$.
This term is diagonal in the nuclear core, specified here by its total angular 
momentum $J_N$ and isospin $T$, with matrix elements that resemble the Fermi 
matrix elements in $\beta$ decay of the core nucleus. Similarly, matrix 
elements of the spin-spin term in Eq.~(\ref{eq:V_YN}) involve the SU(4) 
generator $\sum_j{{\vec s}_{Nj}{\vec t}_{Nj}}$ for the core, connecting core 
states with large Gamow-Teller transition matrix elements. A complete listing 
of these $\Lambda N\leftrightarrow\Sigma N$ Fermi and Gamow-Teller matrix 
elements together with corresponding $\Lambda N$ spin-spin matrix elements 
for the $A=3,4$ $\Lambda$ hypernuclei is given in the first three rows of 
Table~\ref{tab:A=3}, and the resulting binding-energy contributions arising 
from $V_{\Lambda\Sigma}$ are listed in the last two rows, including two-body 
as well as three-body terms. 

\begin{table}[!ht] 
\begin{center}
\caption{Nonvanishing $\Lambda N$ spin-spin matrix elements as well as Fermi 
(F) and Gamow-Teller (GT) nonvanishing matrix elements of $V_{\Lambda\Sigma}$, 
Eq.~(\ref{eq:V_YN}), are listed in the first three rows for 
\lamb{3}{H}($T,J^P$) and \lamb{4}{H}($T,J^P$) $0s_{\Lambda}$ states. 
Estimates of the total $\Lambda\Sigma$ contributions to binding energies, 
using the NSC97e parameter values (\ref{eq:VbarDelta}), are given in MeV 
in the last two rows. Note: $\Delta_{\Lambda\Lambda}$ is positive for 
binding-energy contributions.} 
\begin{tabular}{cccccc} 
\hline 
 & \lamb{3}{H}(0,$\frac{1}{2}^+$) & \lamb{3}{H}(0,$\frac{3}{2}^+$) & 
\lamb{3}{H}(1,$\frac{1}{2}^+$) & \lamb{4}{H}($\frac{1}{2}$,0$^+$) & 
\lamb{4}{H}($\frac{1}{2}$,1$^+$)   \\ 
\hline 
$\Lambda N$ ($\times \Delta_{\Lambda\Lambda}$)& 1 & $-1/2$ &--& 3/4 &$-$1/4 \\ 
F ($\times{\bar V}_{\Lambda\Sigma}$) &--&--& $2\sqrt{2/3}$ & 1 & 1 \\  
GT ($\times\Delta_{\Lambda\Sigma}$) & $\sqrt{3}/2$ &--& $-$1/2 &3/4&$-$1/4 \\ 
$\frac{1}{80}(|{\rm F}|^2 + |{\rm GT}|^2)$ & 0.243 &--& 0.373 &--&-- \\ 
$\frac{1}{80}|({\rm F}+{\rm GT})|^2$ &--&--&--& 0.574 & 0.036 \\ 
\hline
\end{tabular} 
\label{tab:A=3} 
\end{center} 
\end{table} 

The last two columns of the table list matrix elements and binding-energy 
contributions for the $A=4$ states, marked here by \lamb{4}{H}. 
Fermi and Gamow-Teller contributions are added coherently because both 
${\bar V}_{\Lambda\Sigma}$ and $\Delta_{\Lambda\Sigma}$ connect to the same 
and only spin-isospin SU(4) $0s_N0s_{\Sigma}$ intermediate state available. 
The $\Lambda N\leftrightarrow\Sigma N$ transition matrix elements are seen to 
provide about half of the observed 1.1~MeV $0^+_{\rm g.s.}$--$1^+_{\rm exc}$ 
splitting in the $A=4$ hypernuclei, the rest must then be assigned to the 
$\Lambda N$ spin-spin matrix element $\Delta_{\Lambda\Lambda}$. For the $A=3$ 
states, marked here by \lamb{3}{H}, Fermi and Gamow-Teller contributions are 
added incoherently owing to different intermediate states involved in these 
transitions, with binding-energy contributions obtained upon assuming 
implicitly same-size nucleon and hyperon wavefunctions as for $A=4$. Since 
\lamb{3}{H}(0,$\frac{1}{2}^+$) is weakly bound, the actual $A=3$ contributions 
are expected to be somewhat suppressed, with matrix-element suppression factor 
$\eta$ estimated to be about $\eta\approx$0.7--0.8. Even so, given the size 
of both $\Lambda N$ spin-spin and $\Lambda\Sigma$ transition binding-energy 
negative contributions to \lamb{3}{H}(0,$\frac{3}{2}^+$) with respect 
to \lamb{3}{H}(0,$\frac{1}{2}^+$) g.s., it is safe to conclude that 
\lamb{3}{H}(0,$\frac{3}{2}^+$) is unbound.  

Focusing on discussion of \lamb{3}{H}(1,$\frac{1}{2}^+$), particularly 
relative to \lamb{3}{H}(0,$\frac{1}{2}^+$) g.s., we first go to the SU(4) 
limit of nuclear-core dynamics in which the dineutron becomes bound and 
degenerate with the deuteron, and where the difference in $\Lambda$ separation 
energies of \lamb{3}{H}(1,$\frac{1}{2}^+$) and \lamb{3}{H}(0,$\frac{1}{2}^+$) 
according to Table~\ref{tab:A=3} is given (in MeV) by 
\begin{equation} 
\delta B_{\Lambda} \equiv B_{\Lambda}^{T=1}(\;\frac{1}{2}^+) 
- B_{\Lambda}^{T=0}(\;\frac{1}{2}^+) = \eta^2(0.373-0.243) 
- \eta\Delta_{\Lambda\Lambda}. 
\label{eq:DelE} 
\end{equation} 
To maximize this energy difference we neglect the $\Lambda N$ spin-spin 
contribution, thereby letting $\Delta_{\Lambda\Lambda}\to 0$, and compensate 
by doubling the $\Lambda\Sigma$ contribution in order to keep $E(1^+)-E(0^+)
\approx 1.1$~MeV in \lamb{4}{H} intact. For $\eta=1$, expected to be a fair 
approximation in this SU(4) limit, we obtain $\delta B_{\Lambda}^{\rm max}
$=0.26~MeV, and so by charge independence the $\Lambda$ separation energy 
in this hypothetically bound \lamb{3}{n} with respect to the bound dineutron 
core is 0.39$\pm$0.05~MeV. Precisely the same result is obtained if Nijmegen 
model NSC97f $\Lambda\Sigma$ matrix elements from (\ref{eq:VbarDelta}), 
in parentheses there, are used instead. Next, by solving $\Lambda nn$ 
Faddeev equations we fit a $\Lambda N$ spin-independent Yamaguchi separable 
interaction that reproduces $B_{\Lambda}$(\lamb{3}{n})=0.39~MeV, with 
$B(^2{\rm n})$=2.23~MeV as in the deuteron. For a chosen value of 2.5~fm 
for the $\Lambda N$ effective range, this requires a $\Lambda N$ scattering 
length of $-$1.804~fm. For $nn$ interaction we used Yamaguchi separable 
potential determined by the $NN$ $T=0$ low-energy parameters $a_s$=5.4~fm, 
$r_s$=1.75~fm, resulting in $B(^2{\rm n})$=2.23~MeV which equals the deuteron 
binding energy in this SU(4) limit. We then perform a series of $\Lambda nn$ 
Faddeev calculations keeping the $\Lambda N$ interaction as is, but breaking 
SU(4) progressively by varying the $nn$ interaction to reach $a_s$=$-$17.6~fm 
and $r_s$=2.88~fm as appropriate in the real world to the unbound dineutron. 
This is documented in Table~\ref{tab:nn}. 

\begin{table}[!ht] 
\begin{center} 
\caption{Binding energy $B(^2{\rm n})$ (in MeV) of two neutrons in a separable 
Yamaguchi potential specified by scattering length $a_s$ and effective range 
$r_s$ (both in fm) in the $^1S_0$ channel, and $\Lambda$ separation energy 
$B_{\Lambda}$(\lamb{3}{n}) (in MeV) obtained by solving $\Lambda nn$ Faddeev 
equations with a separable Yamaguchi $\Lambda N$ spin-independent interaction 
specified by scattering length $a=-1.804$~fm and effective range $r=2.5$~fm. 
The $B(^2{\rm n})_{\rm approx}$ values are obtained using 
Eq.~(\ref{eq:approx}).} 
\begin{tabular}{ccccc} 
\hline 
$a_s$ & $r_s$ & $B(^2{\rm n})$ & $B(^2{\rm n})_{\rm approx}$ & 
$B_{\Lambda}$(\lamb{3}{n}) \\ 
\hline 
5.4 & 1.75 & 2.23 & 2.24 & 0.39 \\ 
5.4 & 2.25 & 2.79 & 2.87 & 0.27 \\ 
5.4 & 2.881 & 4.98 & -- & 0.16 \\ 
6.0 & 2.881 & 2.86 & 3.20 & 0.11 \\ 
7.0 & 2.881 & 1.64 & 1.68 & 0.06 \\ 
9.0 & 2.881 & 0.80 & 0.80 & 0.01 \\ 
13.0 & 2.881 & 0.32 & 0.32 & 0.003 \\ 
17.612 & 2.881 & 0.16 & 0.16 & -- \\ 
$-$17.612 & 2.881 & -- & -- & -- \\  
\hline 
\end{tabular} 
\label{tab:nn} 
\end{center} 
\end{table} 

The table demonstrates the behavior of the dineutron binding 
energy $B(^2{\rm n})$ and the \lamb{3}{n} binding energy 
$B$(\lamb{3}{n})=$B(^2{\rm n})$+$B_{\Lambda}$(\lamb{3}{n}) upon varying the 
$NN$ low-energy scattering parameters from values given by the $T=0$ $pn$ 
interaction down to the empirical values for the $T=1$ $nn$ interaction. 
This is done in two stages. First, increasing the effective range while 
keeping the scattering length fixed, $B(^2{\rm n})$ increases whereas 
$B_{\Lambda}$(\lamb{3}{n}) steadily decreases.{\footnote{A decrease of 
$B_{\Lambda}$ upon increasing one of the effective ranges in a few-body 
calculation was noted and discussed by Gibson and Lehman \cite{Gibson88}.}} 
In the second stage, while keeping the effective range fixed at its final 
empirical $nn$ value, the scattering length is varied by increasing it and 
then crossing from a large positive value associated with a loosely bound 
dineutron to the empirical large negative value of $a_{nn}$ associated with 
a virtual dineutron. During this stage, $B(^2{\rm n})$ too decreases steadily 
until \lamb{3}{n} is no longer bound. 

With $B_{\Lambda}$(\lamb{3}{n})$\ll B(^2{\rm n})$ holding over the full range 
of variation exhibited in Table~\ref{tab:nn}, it is clear that the behavior 
of $B$(\lamb{3}{n}) follows closely that of $B(^2{\rm n})$. For fairly small 
values of $B(^2{\rm n})$, say $B(^2{\rm n})\lesssim 3$~MeV, $B(^2{\rm n})$ is 
quite accurately reproduced by the effective-range expansion approximation 
\begin{equation} 
B(^2{\rm n})_{\rm approx}=\frac{{\hbar}^2}{M_n r_s^2}
{\left( 1-\sqrt{1-\frac{2r_s}{a_s}} \right)}^2, 
\label{eq:approx} 
\end{equation} 
as shown by comparing the exact and approximate values of $B(^2{\rm n})$ 
listed in the table. 

It is worth noting in Table~\ref{tab:nn} that the dissociation of \lamb{3}{n} 
occurs while the dineutron is still bound, although quite weakly. The final 
result of no \lamb{3}{n} bound state, for a virtual dineutron and $\Lambda N$ 
low-energy scattering parameters listed in the caption to Table~\ref{tab:nn}, 
should come at no surprise given that a considerably larger-size $\Lambda N$ 
scattering length was found to be required in the Faddeev calculations listed 
in Table~\ref{tab:acrit} to bind \lamb{3}{n}. Although a specific value 
of 2.5~fm for the $\Lambda N$ effective range was used in our actual 
demonstration, similar results are obtained for other reasonable choices of 
the $\Lambda N$ effective range.

\section{Discussion and conclusion} 
\label{sec:summ} 

We have shown in this work that the $\Lambda N$ interactions required to bind 
\lamb{3}{n} are inconsistent with the measured $\Lambda p$ scattering cross 
sections at low energies, with \lamb{3}{H}$_{\rm g.s.}$ binding energy, and 
with the $0^+_{\rm g.s.}$--$1^+_{\rm exc}$ excitation energy of the $A=4$ 
$\Lambda$ hypernuclei. Although simple $\Lambda N$ interactions were used to 
simulate the more realistic NSC97 interactions, the consequences of accepting 
a bound \lamb{3}{n} for $\Lambda$ hypernuclear data are sufficiently strong 
that the use of more refined interactions is unlikely to modify any of the 
conclusions reached here. Of the three hypernuclear systems related here to 
\lamb{3}{n}, we attach special significance to the $A=4$ $\Lambda$ hypernuclei 
where only the 1.1~MeV $0^+_{\rm g.s.}$--$1^+_{\rm exc}$ excitation energy 
is involved in our model building. This excitation energy is intimately 
connected to $\Lambda N\leftrightarrow\Sigma N$ coupling effects in the 
$A=4$ hypernuclei \cite{Akaishi00} which have been further incorporated 
and tested successfully to reproduce excitation spectra in $p$-shell 
hypernuclei \cite{Millener12}. We judiciously avoided relying on the absolute 
binding energy of the $0^+_{\rm g.s.}$ of the $A=4$ $\Lambda$ hypernuclei 
because it has not been yet reproduced satisfactorily in few-body 
calculations that use theoretically derived $YN$ potentials, as stressed 
recently by Nogga~\cite{Nogga12}. This difficulty might be associated 
with missing three-body $\Lambda NN$ interaction terms, other than those 
incorporated here by including $\Lambda N\leftrightarrow\Sigma N$ coupling. 

Of the $\Lambda NN$ interactions considered in past hypernuclear 
calculations, those arising from an intermediate $\Sigma(1385)$ 
hyperon resonance \cite{BLN67} are independent of the spin of the 
$\Lambda$ and thus would not affect the $0^+_{\rm g.s.}$--$1^+_{\rm exc}$ 
spin-flip excitation upon which our considerations rest. The spin-isospin 
dependence of the central component of this interaction is given by 
$-({\vec\tau}_1 \cdot {\vec\tau}_2\;{\vec\sigma}_1 \cdot {\vec\sigma}_2)$ 
which assumes the same value $+3$ for both $J^P=\frac{1}{2}^+$ 
states in the $A=3$ hypernuclei. A dispersive $\Lambda NN$ 
repulsive contribution with $\Lambda$ spin dependence given 
by (1+$\frac{1}{3}{\vec\sigma}_{\Lambda}\cdot {\vec S}_{12}$), 
where ${\vec S}_{12}=\frac{1}{2}({\vec\sigma}_1+{\vec\sigma}_2)$, 
was considered in VMC calculations of light hypernuclei \cite{Bodmer88}. 
This gives 1($\frac{1}{3}$) for the $T=1(0),J^P=\frac{1}{2}^+$ $A=3$ states, 
namely more repulsion for \lamb{3}{n} than for \lamb{3}{H}$_{\rm g.s.}$. 
Another form of dispersive $\Lambda NN$ contribution suggested in 
Ref.~\cite{Gal91} depends on spin and isospin through the factor 
$-{\vec\tau}_1 \cdot {\vec\tau}_2 ({\vec\sigma}_1 \cdot {\vec\sigma}_2 + 
{\vec\sigma}_{\Lambda}\cdot {\vec S}_{12})$ which assumes values $+3(-3)$ 
for the $T=1(0)$, $J^P=\frac{1}{2}^+$ states, repulsive for \lamb{3}{n} 
while attractive for \lamb{3}{H}$_{\rm g.s.}$. The latter two dispersive 
$\Lambda NN$ interaction forms were found in Ref.~\cite{Shoeb99} capable of 
accounting for a substantial fraction of the $0^+_{\rm g.s.}$--$1^+_{\rm exc}$ 
excitation in the $A=4$ hypernuclei, but obviously neither of them would 
add attraction to \lamb{3}{n} relative to \lamb{3}{H}$_{\rm g.s.}$. 
This brief survey of three-body $\Lambda NN$ phenomenology offers, therefore, 
no plausible solution of the \lamb{3}{n} puzzle. 

A comment on CSB effects in light $\Lambda$ hypernuclei and whether or not 
CSB might resolve the \lamb{3}{n} puzzle is in order before concluding the 
present study. For the known $T=\frac{1}{2}$ isodoublet of $A=4$ hypernuclear 
$0^+_{\rm g.s.}$ levels $\Delta B_{\Lambda}^{\rm exp}(A=4)$$\equiv$$
B_{\Lambda}$(\lamb{4}{He})$-$$B_{\Lambda}$(\lamb{4}{H})=0.35$\pm$0.04~MeV 
\cite{Davis05} is exceptionally large and defies explanation in modern $YN$ 
interaction models, see Table~9 in Ref.~\cite{Nogga12} where the recently 
constructed NLO chiral $YN$ interactions~\cite{Haiden13} are shown to yield 
only $\Delta B_{\Lambda}^{\rm calc}(A=4)\approx 50$~keV. This $\Delta B_{
\Lambda}(A=4)$ arises largely from kinetic energies depending on which charged 
$\Sigma$ hyperon mass is used. The same CSB effect will result in smaller 
$B_{\Lambda}$(\lamb{3}{n}) values relative to those calculated, as done here, 
using a charge symmetric calculation. Therefore, CSB contributions are also 
unlikely to resolve the \lamb{3}{n} puzzle. 

How does one then explain the HypHI $t+\pi^-$ signal which is naturally 
assigned to the two-body weak decay \lamb{3}{n}$\to t+\pi^-$? This problem 
is aggravated by a similar one addressing a $d+\pi^-$ signal, also observed 
in the HypHI experiment, the most straightforward assignment of which would 
be due to the two-body weak decay of a bound $\Lambda n$ system: 
\lamb{2}{n}$\to d+\pi^-$. No plausible solution has been offered to these 
puzzles and more work on other possible origins of $d+\pi^-$ and $t+\pi^-$ 
signals is called for. 

\section*{Acknowledgements} 

We thank Nir Barnea, Daniel Gazda, Emiko Hiyama, Ji\v{r}\'{i} Mare\v{s} 
and Jean-Marc Richard for useful discussions. A.G. acknowledges support 
by the EU initiative FP7, HadronPhysics3, under the SPHERE and LEANNIS 
cooperation programs. H.G. is supported in part by COFAA-IPN (M\'exico).


\begin{thebibliography}{99} 

\bibitem{Davis05} D.H.~Davis, Nucl. Phys. A 754 (2005) 3c, and references 
listed therein. 

\bibitem{DD59} B.W.~Downs, R.H.~Dalitz, Phys. Rev. 114 (1959) 593. 

\bibitem{Gar87} H.~Garcilazo, J. Phys. G 13 (1987) L63. 

\bibitem{MKGS95} K.~Miyagawa, H.~Kamada, W.~Gl\"{o}ckle, V.~Stoks, 
Phys. Rev. C 51 (1995) 2905. 

\bibitem{Hiyama14} E.~Hiyama, S.~Ohnishi, B.F.~Gibson, Th.A.~Rijken, 
Phys. Rev. C 89 (2014) 061302(R).  

\bibitem{Gar07a} H.~Garcilazo, T.~Fern\'{a}ndez-Caram\'{e}s, A.~Valcarce, 
Phys. Rev. C 75 (2007) 034002. 

\bibitem{Gar07b} H.~Garcilazo, A.~Valcarce, T.~Fern\'{a}ndez-Caram\'{e}s, 
Phys. Rev. C 76 (2007) 034001. 

\bibitem{Gar14} H.~Garcilazo, A.~Valcarce, Phys. Rev. C 89 (2014) 057001. 

\bibitem{Gazda14} D.~Gazda, private communication (May 2014). 

\bibitem{Haiden13} J.~Haidenbauer, S.~Petschauer, N.~Kaiser,
U.-G.~Mei{\ss}ner, A.~Nogga, W.~Weise, Nucl. Phys. A 915 (2013) 24,
and references listed therein. 

\bibitem{JMR14} J.M.~Richard, Q.~Wang, Q.~Zhao, arXiv:1404.3473(nucl-th). 

\bibitem{Rap13} C.~Rappold, et al. (HypHI Collaboration), Phys. Rev. C 88 
(2013) 041001(R). 

\bibitem{Alex68} G.~Alexander, U.~Karshon, et al., Phys. Rev. 173 (1968) 1452. 

\bibitem{Rijken10} Th.A.~Rijken, M.M.~Nagels, Y.~Yamamoto, Prog. Theor. Phys. 
185 (2010) 14, and references listed therein. 

\bibitem{Nogga12} A.~Nogga, Nucl. Phys. A 914 (2013) 140, and references 
listed therein. 


\bibitem{Akaishi00} Y.~Akaishi, T.~Harada, S.~Shinmura, K.S.~Myint, 
Phys. Rev. Lett. 84 (2000) 3539. 

\bibitem{Millener12} D.J.~Millener, Nucl. Phys. A 881 (2012) 298, 
and references listed therein. 

\bibitem{Galmil13} A.~Gal, D.J.~Millener, Phys. Lett. B 725 (2013) 445. 

\bibitem{Millener08} D.J.~Millener, Nucl. Phys. A 804 (2008) 84. 

\bibitem{Gibson88} B.F.~Gibson, D.R.~Lehman, Phys. Rev. C 37 (1988) 679, 
and references listed therein. 

\bibitem{BLN67} R.~Bhaduri, B.~Loiseau, Y.~Nogami, Ann. Phys. 44 (1967) 57; 
A.~Gal, J.M.~Soper, R.H.~Dalitz, Ann. Phys. 63 (1971) 53.  

\bibitem{Bodmer88} A.R.~Bodmer, Q.N.~Usmani, Nucl. Phys. A 477 (1988) 621. 

\bibitem{Gal91} A.~Gal, in {\it Plots, Quarks and Strange Particles}, 
Eds. I.J.R.~Aitchison, C.H.~Llewellyn Smith, J.E.~Paton (WS, Singapore, 1991) 
pp.~146-158. 

\bibitem{Shoeb99} M.~Shoeb, N.~Neelofer, Q.N.~Usmani, M.Z.~Rahman Khan, 
Phys. Rev. C 59 (1999) 2807. 




\end{thebibliography}
\end{document}